\documentclass[sigconf,natbib=true]{acmart}

\usepackage{multirow}
\newcommand{\bfs}{\textsf{BFS}}
\newcommand{\qoracle}{\textsf{QOracle}}
\newcommand{\qfirst}{\textsf{QFirst}}
\newcommand{\qmin}{\textsf{QMin}}

\copyrightyear{2025}
\acmYear{2025}
\setcopyright{cc}
\setcctype{by}
\acmConference[ICTIR '25]{Proceedings of the 2025 International ACM SIGIR Conference on Innovative Concepts and Theories in Information Retrieval (ICTIR)}{July 18, 2025}{Padua, Italy}
\acmBooktitle{Proceedings of the 2025 International ACM SIGIR Conference on Innovative Concepts and Theories in Information Retrieval (ICTIR) (ICTIR '25), July 18, 2025, Padua, Italy}\acmDOI{10.1145/3731120.3744597}
\acmISBN{979-8-4007-1861-8/2025/07}

\begin{document}


\title{Neural Prioritisation for Web Crawling}


\author{Francesca Pezzuti}
\email{francesca.pezzuti@phd.unipi.it}
\affiliation{%
  \institution{University of Pisa}
  \city{Pisa}
  \country{Italy}
}

\author{Sean MacAvaney}
\email{sean.macavaney@glasgow.ac.uk}
\affiliation{%
 \institution{University of Glasgow}
 \city{Glasgow}
 \country{UK}}

\author{Nicola Tonellotto}
\email{nicola.tonellotto@unipi.it}
\affiliation{%
  \institution{University of Pisa}
  \city{Pisa}
  \country{Italy}}
  
\renewcommand{\shortauthors}{Pezzuti et al.}

\begin{abstract}
Given the vast scale of the Web, crawling prioritisation techniques based on link graph traversal, popularity, link analysis, and textual content are frequently applied to surface documents that are most likely to be valuable.
While existing techniques are effective for keyword-based search, both retrieval methods and user search behaviours are shifting from keyword-based matching to natural language semantic matching.
The remarkable success of applying semantic matching and quality signals during ranking leads us to hypothesize that crawling could be improved by prioritizing Web pages with high semantic quality.
To investigate this, we propose a semantic quality-driven prioritisation technique to enhance the effectiveness of crawling and align the crawler behaviour with recent shift towards natural language search.
We embed semantic understanding directly into the crawling process --- leveraging recent neural semantic quality estimators to prioritise the crawling frontier --- with the goal of surfacing content that is semantically rich and valuable for modern search needs.
Our experiments on the English subset of ClueWeb22-B and the Researchy Questions query set show that, compared to existing crawling techniques, neural crawling policies significantly improve harvest rate, maxNDCG, and search effectiveness during the early stages of crawling.
Meanwhile, crawlers based on our proposed neural policies maintain comparable search performance on keyword queries from the MS MARCO Web Search query set.
While this work does not propose a definitive and complete solution, it presents a forward-looking perspective on Web crawling and opens the door to a new line of research on leveraging semantic analysis to effectively align crawlers with the ongoing shift toward natural language search.
\end{abstract}



\maketitle
\renewcommand{\shortauthors}{Francesca Pezzuti, Sean MacAvaney, and Nicola Tonellotto} 

\section{Introduction}
    The effectiveness of search engines heavily depends on the indexed corpus; if the corpus is incomplete, outdated, or filled with low-quality pages, search results could be irrelevant or inaccurate to users~\cite{lewandowski2008retrievalqual}.
    Search engines typically employ crawling techniques to automate the discovery of large-scale corpora of Web documents.
    Specifically, a \textit{crawler} is a program that systematically traverses the Web and downloads Web pages to build and keep up-to-date a corpus for Web search. The ability of the crawler to prioritise high-quality pages is crucial for ensuring that the search engine provides accurate, relevant, and up-to-date information to users~\cite{fetterly2009impactpolicy}.
    
   The crawler maintains a priority queue of URLs of pages to visit, called a frontier. 
    It continuously downloads the page at the next URL from the frontier, extracts its outgoing links, and prioritizes them in the frontier.
    Traditional graph traversal algorithms, such as \textit{Breadth-First Search} (BFS) or \textit{Depth-First Search} (DFS), can be used for traversal~\cite{najork2001breadth,debra1994fishsearch}.
    While both BFS and DFS do not utilise any heuristic to decide which URL to process next, \textit{Best-First} policies (BF) are designed to select the most \textit{valuable} URL from the frontier according to some heuristic (e.g., click-through rates~\cite{ostroumova2014ctr}, PageRank~\cite{page1999pagerank}, or textual content~\cite{najork2001breadth, ostroumova2014ctr}). However, each heuristic introduces its own limitations.
    For instance, popularity-based BF crawlers prioritise highly clicked pages over less visited ones, for example based on a Click-Through Rate (CTR) predictor trained on historical CTR information to estimate the CTR value of newly discovered pages~\cite{ostroumova2014ctr}. However, these popularity-based approaches require massive amounts of historical click data that may not always be accessible, tend to neglect newly created pages~\cite{pandey2005ctrproblem}, and are unreliable for ephemerally popular pages, unless discounted in time~\cite{ostroumova2014ctr}.
    On the other hand, PageRank, --- the most notable example of connectivity-based quality heuristic --- assigns higher priority to well-linked pages, i.e., pages linked by many important pages. However, it requires resource-intensive computations~\cite{menczer2004topicalcrawl}, requires storing the full Web graph, is biased towards old pages~\cite{baeza2002webdynamics}, and is a global measure~\cite{menczer2004topicalcrawl} that is known to yield inaccurate estimates when applied to partial Web graphs~\cite{boldi2004paradox,holzmann2019pagerankdeviations}.
    Content-based quality-estimation heuristics have mainly been used in query-driven, topic-specific crawling processes by focused crawlers~\cite{chakrabarti1999focusedcrawl,menczer2004topicalcrawl,tyagi2023focusedcrawl}. Crawlers of this kind typically rely on machine-learning classifiers~\cite{chakrabarti1999focusedcrawl,pant2006harvstrate}, on reinforcement learning~\cite{mccallum2000RLfocusedcrawl}, on ontologies~\cite{ehrig2003ontologycrawl}, on the vector space model~\cite{salton1975} by prioritising pages with high textual similarity w.r.t. a query or a set of relevant documents selected to drive the crawl~\cite{cho1998tfidf, hersovici1998sharksearch,pant2003symbiosis}, or on keyword matching w.r.t. topic keywords~\cite{debra1994fishsearch}.
    However, in all these focused crawlers, the crawling process  is query- or topic-driven as their goal is to target documents relevant to specific queries or topics. However, these approaches are beyond the scope of our work as we are interested in general-purpose, query-agnostic crawling for generic Web search systems.
   
    Although all aforementioned prioritisation policies work well with classical keyword-based queries, they either (i) do not leverage the textual content of pages, (ii) use it for keyword matching w.r.t. topic keywords, or (iii) use it for query-driven focused-crawling, relying on relevance signals based on term frequency and inverse document frequency, ignoring the semantics of texts.
    With recent advancements in contextualised Large Language Models (LLMs), search on the Web is becoming more conversational and users are increasingly submitting question-based, complex queries rather than simple keywords~\cite{trippas2024queriesLLMs, guy2016voicequeries}. Consequently, many search applications including question answering systems, LLM-based assistants and mobile voice search, shifted their focus from short keyword queries to natural language ones~\cite{white2024searchAI,Tajmir2024keywordsearch,guy2016voicequeries,chen2021queryreformulation,xie2023ttiprompt,SauchukTHTS22}.
    
    In this paper, we argue that crawlers should adapt to the shift towards natural language queries.
    We hypothesise that prioritising Web pages based on quality estimates considering both syntactical structure and text semantics can improve the ability of crawlers to surface documents valuable for search tasks, and in particular, pages relevant to natural language queries.
    To this end, we propose to prioritise high-quality pages during crawling by leveraging pre-trained LLMs, which have a semantic understanding of text.
    Our ultimate goal is to devise a crawling policy that improves effectiveness early in the crawling process --- particularly for natural language queries --- and better aligns with recent search trends.
    
    Indeed, researchers have already begun exploring the use of LLMs for prioritising the frontier during crawling. 
    In fact,~\citet{yu2025craw4llm} recently proposed \textsc{CRAW4LLM}, a crawling approach that leverages a LLM to effectively target data of high-quality for LLM pre-training.
    Similarly,~\citet{pezzuti2025qualscor} showed the potential benefit of using a LLM to drive the crawling process by prioritising semantically high-quality Web pages.
    In particular, their early work explored how an oracle crawler --- inapplicable in real-world crawling scenarios, but providing a theoretical upper-bound on the achievable benefit --- effectively increases the downstream recall of a search system.
    Notably, they also observed that documents of similar semantic quality tend to link to one another, suggesting the possibility of approximating the semantic quality of one page with that of its neighbours.
    However, to the best of our knowledge, this last idea remains unexplored.

    To fill this gap, in this paper, we build on the approach by~\citet{pezzuti2025qualscor}, and by relying on the assumption that Web pages with similar quality are likely to link to each other, we propose two novel neural crawling policies based on quality propagation --- by exploiting the inlinking neighbourhood of Web pages ---, and based on neural quality estimators~\cite{chang2024qpruning}.
    
    Notably, these neural quality estimators are LLMs trained to predict whether a document is likely to be relevant to any query submitted to the search engine, in a query-independent way, and independently of other documents.
    Interestingly, being performed independently of other pages, quality estimation using neural quality estimators can be performed on-the-fly, or for example during the active waits of the crawling process caused by network latency and by politeness constraints~\cite{kolobov2019politeness}.
    This independency, also makes our approach easily scalable and parallelisable for large-scale Web crawling scenarios.
    Moreover, while our approach focuses on prioritising Web pages based on the semantic quality computed from the plaintext of Web pages, it can be easily extended to incorporate additional signals, such as language, metadata, and anchor text without requiring significant changes to the crawler component or the search system.
    This flexibility opens up numerous possibilities for further enhancing Web crawling processes.
    Furthermore, the underlying neural quality estimator  can be (re-)fine-tuned to quickly adapt to emerging search trends and to new user search behaviours, ensuring continuous alignment with dynamic definitions of semantic quality and search relevance.
    
    While in literature there is strong evidence for the potential of utilising LLMs for prioritising the frontier during crawling, both for targeting high-quality LLM pre-training data and for generic search tasks, the alignment of crawlers with the recent trend towards natural language search is yet to be explored.
    To investigate this, we compare the crawling effectiveness and the effectiveness of a multi-stage ranking pipeline on keyword queries and natural language queries, for corpora of documents built by a classical BFS crawler, and by crawlers based on neural prioritisation.
    
    In particular, to assess the impact of our crawling policies on the ability of a crawler to find valuable pages early, we compute the \textit{harvest rate} (HR) considering as target all documents relevant to at least a query~\cite{chakrabarti1999focusedcrawl,pant2006harvstrate}. Following prior research, to assess the impact on the potential search effectiveness, we compute \textit{maxNDCG}, i.e., the nDCG achievable by the ideal ranker~\cite{fetterly2009impactpolicy}. Finally, to evaluate search effectiveness, we measure nDCG@10 for a \textit{BM25$\gg$MonoELECTRA}
    multi-stage ranking pipeline. All the metrics are computed dynamically, i.e., while the crawler downloads Web pages.
    We are particularly interested in maximising early crawling and search effectiveness, since relevant pages are generally more dynamic than irrelevant ones~\cite{elsas2010temporaldynamics}.
   
    Our reproducible experiments on the English subset of ClueWeb22-B~\cite{overwijk2022clueweb} show that our proposed neural crawling policies can significantly improve HR (up to $+149\%$), maxNDCG (up to $+152\%$), and nDCG@10 (up to $+139\%$) for natural language queries from Researchy Questions~\cite{rosset2024rqdataset}, when compared to BFS, across the entire crawling process.
    Meanwhile, for keyword queries from MS MARCO Web Search~\cite{chen2024msmarco}, they can consistently improve HR (up to $+20\%$), substantially improve maxNDCG (up to $+20\%$) during early crawl stages, and remain competitive in terms of nDCG@10 compared to BFS.
    Moreover, we also show that we can markedly improve efficiency in terms of mean speedup throughout a crawl both on natural language queries (up to $1.6\times$) and keyword queries (up to $1.1\times$).
    
    Because of the natural small-scale of our crawling experiments (we crawl the English subset of  head pages from ClueWeb22-B), PageRank scores would be unreliable page importance estimates~\cite{boldi2004paradox,holzmann2019pagerankdeviations}. As such, we do not include PageRank as a baseline. Instead, we rely on BFS, which is a well-established strong baseline for graphs limited in size that performs a level-by-level visit of the Web graph and implicitly favours  pages of high graph centrality, approximating PageRank behaviour~\cite{najork2001breadth}.
    
    However, while our experiments are limited in scale, they are fully reproducible and may provide an important step toward understanding the potential of integrating LLMs into Web crawlers.
    
    In summary, we challenged the traditional assumptions on \textit{how to crawl} by rethinking how crawlers prioritise the content. By shifting from the exploitation of metrics that overlook the semantic quality of Web pages --- such as links and popularity ---, to prioritising semantic quality itself, we unlock many major advantages. These include improved scalability, flexibility, and effectiveness, as well as better adaptability to evolving search behaviours.

\section{Neural Crawling Policies}

    We now describe our framework for neural crawlers based on neural quality estimators (Section~\ref{ssec:framework}), and their complexity and scalability (Section~\ref{ssec:scalability}).

\subsection{Proposed Framework}\label{ssec:framework}
    Let $p$ denote a Web page, and $\mathcal{O}_p$ denote the set of outlinks from $p$ (i.e., pages that $p$ links to).
    Let $\mathcal{F}$ denote the \textit{frontier}, which stores the URLs of the pages yet to be crawled.
    The frontier is composed of $(u, P_u)$ pairs, where $u$ is a URL and $P_u$ is its priority.
    A \textit{crawling policy} is composed by: (i) a \textit{priority-assignment function} $P: u \mapsto P_u$, typically based on heuristics such as PageRank, (ii) an \textit{update policy}, that defines how priorities are updated when discovering a new link to a page whose URL is already in the frontier, and (iii) a \textit{selection policy}, that decides which page to crawl next, according to its priority.
    For instance, the well-known Breadth-First Search (BFS) crawling policy uses a constant priority-assignment function, a First-In-First-Out selection policy, and an update policy that does not change priorities upon rediscovery. In contrast, the Best-First (BF) policy employs a priority-assignment function based on a quality-estimation heuristic like PageRank, a maximum priority selection policy, and often more advanced update policies.  
    Building on the last approach,  we propose two neural BF crawling policies 
    leveraging an LLM-based quality-estimation heuristic to prioritise Web pages with high semantic quality during the crawling process, and we compare them with the oracle neural policy in~\cite{pezzuti2025qualscor}.
    
    At the core of our proposed neural BF crawling policies is the use of a LLM-based heuristic function $M_\theta:p\mapsto\mathbb{R}$, parametrised by $\theta$ and optimised to distinguish high semantic quality pages from low-quality ones. 
    In particular, we aim at exploiting this neural heuristic in the priority-estimation function, ideally using $P: u \mapsto M_\theta(p)$ to prioritise a page $p$ whose URL is $u$.
    However, in real-world crawling settings we cannot access the textual content of a Web page before its download.
    Indeed, using the ideal semantic quality as the priority when enqueueing pages into $\mathcal{F}$ is only feasible in a theoretical scenario where an oracle function has access to a page text prior its download.  We refer to this oracle-based crawling policy as \qoracle, and we use it as an upper-bound on the performance achievable by our practical neural crawling policies.
    
    In the absence of an oracle, we can reasonably assume that the quality of a page is related to the quality of the pages it is connected to. Indeed, existing literature shows that the quality of a page is positively correlated with that of its linking neighbours.
    If this relationship holds, we can effectively propagate quality via link structure by using as a proxy estimate of the quality of a page, that of one of its ancestors (i.e., the page that linked to it).
    Hence, in our first neural crawling policy, referred to as \qfirst, when processing a page $p$ and encountering the outgoing URL $\tilde u$ of a new page $\tilde p \in \mathcal{O}_{p}$ for the first time, we insert $\tilde u$ into $\mathcal{F}$ with priority $P_{\tilde u} =M_\theta (p)$ and we never update it. 
    
    In our second neural crawling policy, referred to as \qmin, we make the additional assumption that if a page is linked to a low-quality page, it is highly unlikely to be of high-quality. If this holds, we could postpone the crawling of low-quality pages by decreasing their priority whenever a link from a low-quality ancestor is discovered. In doing so, we aim to boost the prioritisation high-quality Web pages while deprioritising low-quality ones.
    To implement the \qmin\ policy, when processing a page $p$ and re-encountering an already enqueued URL $\tilde u\in \mathcal{F}$ of a page $\tilde{p} \in \mathcal{O}_p$, we update its priority as the minimum between the current priority and the quality of the new ancestor $p$, i.e., $P_{\tilde{u}} \gets \min \left \{ P_{\tilde{u}}, M_\theta (p)\right \}$.
    
    We chose not to propose a \textsc{QMax} policy, as our preliminary experiments, consistent with findings by~\citet{chang2024qpruning}, and by~\citet{pezzuti2025qualscor}, confirmed that neural estimators better identify low-quality pages than high-quality ones.

\subsection{Complexity \& Scalability}\label{ssec:scalability}
    After having discussed the proposed neural crawling policies based on neural quality estimators, we now analyse their computational implications in terms of time complexity, space complexity and scalability for large-scale Web crawling deployments.
    
    The theoretical time complexity of estimating the quality of a page involves a forward pass through the LLM and depends on its length in terms of tokens $t$, hence it is $O(t^2)$ due to the self-attention layers. In practice, however, the average cost per page can be considered constant because the average token count is stable and bounded when considering large numbers of pages. More importantly, quality estimation performed with neural quality estimators is performed independently on each page, without relying on global corpus information.
    In summary, the time complexity of estimating the quality of $n$ pages is $O(n)$.
    Our approach supports fine-grained, one-the-fly quality estimation, enabling precise frontier prioritisation: the quality of a page can be computed immediately and reliably after its download, and remains valid once computed.
    Alternatively, to balance immediacy and efficiency, one can opt for delayed batched estimation rather than on-the-fly single-page quality estimation.
    Given that the frontier queue grows rapidly in time, the delay introduced by batching and by LLM inference is expected to be negligible.
    Moreover, \citet{chang2024qpruning}, showed that even lightweight language models (such as 4-layer transformers) can be highly effective at estimating semantic quality. Thus, in practice, the inference latency is unlikely to become a bottleneck, further supporting the applicability of our approach.
    
    Conversely, crawling approaches based on PageRank or its variants require the periodic re-computation of scores over the entire portion of discovered Web graph.
    These computations involve expensive operations and have a time complexity $O(n^2)$. 
    As the number of discovered pages grows rapidly in time, the time complexity is not constant in time.
    Moreover, crawlers based on these methods rely on coarse-grained scores computed on sub-graphs, which are poor proxies for the actual authoritativeness of the pages~\cite{boldi2004paradox,holzmann2019pagerankdeviations}. Additionally, when inserting into the frontier newly discovered pages, these crawlers assign random or minimum priorities as scoring with PageRank-like methods is performed only at fixed intervals. Both these issues contribute to suboptimal frontier prioritisation.

    Beyond computational advantages, our approach offers significant benefits in terms of space consumption. Indeed, methods based on link-structure require storing the full Web graph and have space complexity $O(n+e)$ where $e$ is the number of edges, $n$ is the number of nodes, and $e \gg n$. In contrast, our strategy only needs to store a real-valued score per page. As a result,  its space complexity is $O(1)$ per-page, and $O(n)$ in aggregate for $n$ pages.

    From a scalability perspective, neural quality estimation has an \textit{embarrassingly parallelisable} nature. Indeed, since each Web page can be scored independently from the others solely based on its content, quality estimation can be distributed across multiple machines without any inter-process communication overhead or synchronisation. 
    Overall, our approach introduces a constant per-page computational overhead at frontier-prioritisation time, a constant per-page memory overhead, and is highly suitable for scalable Web crawling applications. 
    
    Lastly, in our approach there are no hyper-parameters such as PageRank's iteration threshold and damping factor, whose poor tuning may hamper accuracy~\cite{bianchini2005pagerank}.  This absence of hyper-parameters makes the quality estimation process based on neural quality estimators easily and readily deployable in crawling systems.
 \begin{figure}
        \includegraphics[width=\linewidth]{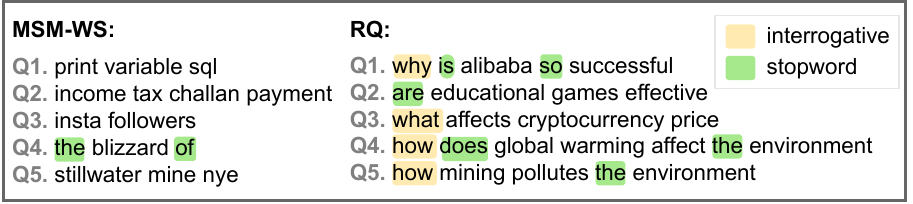}
        \caption{Random sample of $5$ queries from MSM-WS and $5$ queries from RQ.}
        \label{fig:queries_sample}
        \Description[MSM-WS contains keyword queries while RQ contains natural language queries]{A random sample of $5$ queries shows that MSM-WS contains short keyword queries while RQ contains long natural language queries.}
\end{figure}

    \begin{table}
     \centering
      \caption{Sizes of MSM-WS and RQ after filtering.}
     \label{tab:querysets}
     \begin{tabular}{@{}cccccc@{}}
        \toprule
          \bf Query set & \bf $\vert \mathcal{Q} \vert$ & \bf $ \vert \mathcal{R}_{\mathcal{D}} \vert $ & \bf  Qlen & \bf Interrogatives \\
          \midrule
          MSM-WS & $896$ & $894$ & $3.8 \pm 2.3$ & $0.06 \pm 0.25$ \\
          RQ & $2,122$ & $1,973$ & $7.6 \pm 2.8$ & $0.86 \pm 0.41$ \\
        \bottomrule
     \end{tabular}
 \end{table}

\begin{figure*}[t!]
    \includegraphics[width=1\linewidth]{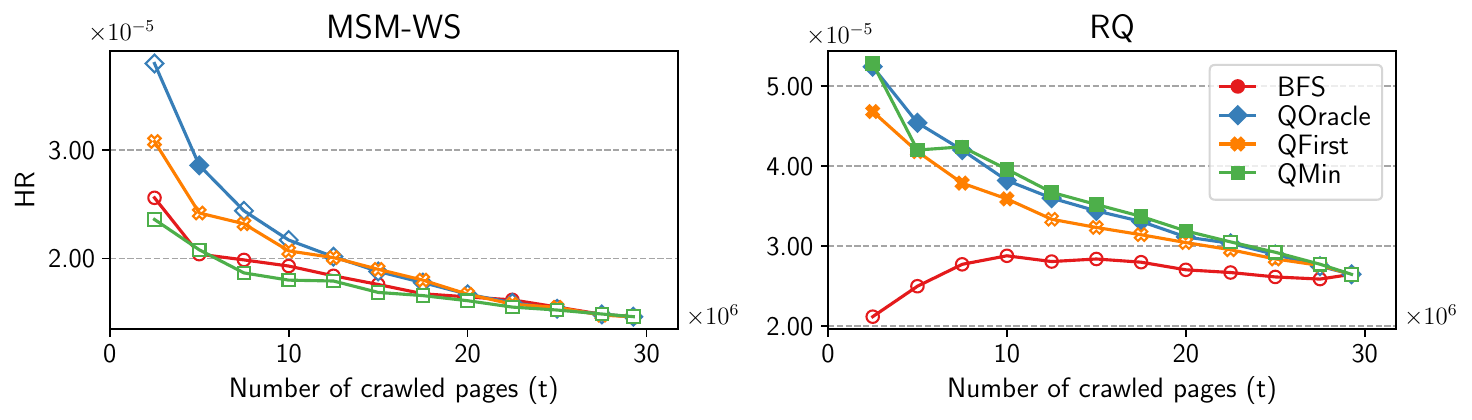}
    \caption{Assessment of neural crawling policies and BFS in terms of Harvest Rate. MSM-WS (left) and RQ (right). Statistically significant differences w.r.t. BFS are denoted with \textit{filled circles} ($p=0.01$).}
    \label{fig:hr-all}
    \Description[Neural policies yield to an increase in HR w.r.t. BFS]{Both for MSM-WS and RQ, our neural crawling policies are significantly more effective than BFS in terms of HR, especially during the first iterations of the crawl.}
\end{figure*}

\section{Experimental Setting}

    We conduct experiments to answer these research questions:
        \begin{description}
            \item[\bf RQ1] Does prioritising pages with high semantic quality during the crawl, increase the effectiveness of crawling?
            \item[\bf RQ2] Does prioritising pages with high semantic quality during the crawl, increase the effectiveness of a retrieval system?
            \item[\bf RQ3] What are the trade-offs in terms of time efficiency and effectiveness of semantic quality crawling?
        \end{description}

\subsection{Web Corpus}
    To ensure replicability and reproducibility of our experiments, we perform a single-threaded simulation of the crawling process using a publicly available dataset. Specifically, we crawl the subset of English pages of ClueWeb22-B (CW22B-eng), that in turn is the subset of $87M$ head Web pages from ClueWeb22 (CW22)~\cite{overwijk2022clueweb}, a recently released corpus built by crawling the Web with a commercial search engine.
    This approach is aligned with real-world recrawling practices, where crawlers typically avoid frequent recrawling of noisy/spam pages, focusing on potentially high-quality pages~\cite{olston2008recrawl}. 
    In our simulations, Web pages are crawled sequentially, and we assume a constant per-page crawling time.
    In real-world scenarios, crawlers iteratively download Web pages and stop after periods of duration $T$ to allow retrievers to update their index.
    To simulate the iterative process, we measure effectiveness every $T=2.5M$ crawled pages.
    We start to crawl from $100k$ randomly selected seed URLs, and we reach a total of $29M$ pages.

\subsection{Queries}
    We use MS MARCO Web Search (MSM-WS)~\cite{chen2024msmarco} and Researchy Questions (RQ)~\cite{rosset2024rqdataset} query sets, both generated from the logs of commercial search engines. 
    MSM-WS contains queries reflecting a real query distribution and relevance labels over CW22 extracted from a real click-log, with explicit relevance assessments.
    The latter contains multi-perspective, non-factoid English queries, and a click distribution over CW22. For each query in RQ, we considered the most clicked page as relevant.
    Since we work with a subset of CW22B, but both query sets are related to CW22, we excluded queries without relevant results in our CW22B-eng subset.  
    Figure~\ref{fig:queries_sample} shows five randomly sampled queries from these two query sets. From this sample we observe that queries from MSM-WS tend to be simple and keyword-based, whereas queries from RQ are more descriptive and expressed in natural language. 
    Details about our filtered versions of MSM-WS and RQ are available in Table~\ref{tab:querysets}. 
    As reported, MSM-WS contains shorter queries compared to RQ, with a narrow scope and mainly keyword-based. Moreover MSM-WS queries contain fewer interrogatives like "how" and "why", while RQ queries have a broader scope and are similar to natural language.
    
\subsection{Neural Quality Estimation}
    In all experiments, we use the \textit{QT5-Small} neural quality estimator by~\citet{chang2024qpruning}. We trained a QT5-Small model using a modified version of the training script provided by the original authors, adjusted to work with the CW22 dataset. As positive quality labels, we sampled from the 9.1M documents with positive relevance labels from the MSM-WS training set, treating all others as negatives. We found the model converged after 1.6M training instances, representing only a fraction of the available training data.
 
\subsection{Effectiveness Metrics}
    There exist several approaches for comparing the effectiveness of crawling policies~\cite{menczer2001crawleval, fetterly2009impactpolicy}. Harvest Rate (HR) is a widely used effectiveness metric for crawling~\cite{chakrabarti1999focusedcrawl,pant2006harvstrate}. Let $\mathcal{R}^{\mathcal{Q}}$ denote the set of all pages relevant to at least a query $q$ in a query set $\mathcal{Q}$. At time $t$, for the query set $\mathcal{Q}$, the harvest rate $HR(\mathcal{Q}, t)$ is computed as:
    \begin{equation*}
    HR(\mathcal{Q}, t) = \frac{\vert \mathcal{R}^{\mathcal{Q}}_t \vert}{t},
    \end{equation*}
    where $\mathcal{R}^{\mathcal{Q}}_t$ is the subset of relevant pages crawled up to time $t$.
    This metric measures the crawl ability to maximise the number of crawled relevant pages while minimising that of irrelevant ones.
    Another notable crawling effectiveness metric is maxNDCG~\cite{fetterly2009impactpolicy}, that is the highest nDCG achievable based on the number of crawled relevant pages.
    For a query $q \in \mathcal{Q}$, $maxNDCG(q, t)$ is computed as: 
    \begin{equation*}
        maxNDCG(q, t) = \sum_{i=1}^{\vert \mathcal{R}^q_t \vert} \frac{1}{\log_2(i+1)},
    \end{equation*}
    where $\mathcal{R}^q_t$ is the set of pages relevant to $q$, crawled up to time~$t$. 
    Following prior research~\cite{fetterly2009ndcg10}, we also validate the impact of our neural crawling policies on downstream retrieval effectiveness. However retrieval effectiveness depends not only on the crawler, but also on the retriever.
    To address this, we periodically measure nDCG@10 for a multi-stage IR pipeline, reranking the top 100 pages retrieved by BM25 using MonoELECTRA\footnote{\url{https://huggingface.co/crystina-z/monoELECTRA_LCE_nneg31}}~\cite{clark2020electra}, a strong cross-encoder model trained with hard negatives for ranking tasks.

\subsection{Efficiency metrics}
    Given the large scale of the Web, another important aspect to consider when evaluating crawlers is their \textit{efficiency} both in collecting the most valuable Web pages, as early as possible during the crawl, and in limiting resource waste, e.g., storage usage and bandwidth usage.
    Efficiency is crucial both in terms of resources and time, because i) crawlers typically are constrained in storage and bandwidth, and should therefore minimise the number of downloads of low-value pages from the Web to save space and network resources, and ii) the crawling process is typically subject to a fixed time budget, since traversing and storing the entire Web is infeasible. 
    
    As already mentioned, we assume a constant per-page crawling time, and we consider the time required to crawl a single page as time unit; under these assumptions, resource efficiency and time efficiency are both inversely proportional to the number of low-valuable pages downloaded.
    Consequently, we propose to compare the efficiency of a crawling approach $A$ w.r.t. a baseline $B$, by measuring how faster $A$ collects a fixed number of relevant pages w.r.t. $B$ --- or, equivalently, how many more pages than $A$ the baseline $B$ must download to collect the same number of relevant pages.
    To do so, let $\tau_A(n)$ and $\tau_B(n)$ denote the time taken by the approaches $A$ and $B$, respectively, to crawl $n$ pages relevant to at least a query in a given query set $\mathcal{Q}$.
    We define the \textit{speedup} at $n$ of $A$ w.r.t. $B$, denoted with $s_{A,B}(n)$, as:
    \begin{equation*}
        s_{A,B}(n) = \frac{\tau_B(n)}{\tau_A(n)},
    \end{equation*}
    where $n$ is the target number of relevant pages.
    
    A speedup $s_{A,B}(n)>1$ indicates that the crawler $A$ is $s_{A,B}(n)$-times faster than the baseline $B$ at collecting $n$ relevant Web pages.
    Conversely, when $s_{A,B}(n)<1$, the crawler $A$ is less efficient than the baseline $B$ since it must crawl more pages than $B$ to collect the same number of relevant ones.
    In summary, our speedup measures how more efficient is a crawling approach compared to a baseline in collecting $n$ valuable pages.
    Lastly, we define the \textit{mean speedup} as the average of the speedups over a complete crawl.
    This aggregate measure quantifies the overall advantage (or disadvantage) of the crawling approach $A$ compared to the baseline $B$ in terms of speedup throughout the crawl --- thus, it measures the overall efficiency of a crawling approach w.r.t. a baseline.

\subsection{Baseline} 
    As already mentioned, we compare our proposed policies against BFS, the simplest yet effective general-purpose policy. Motivated by findings from~\citet{boldi2004paradox,holzmann2019pagerankdeviations}, and~\citet{najork2001breadth}, which showed that on small graphs PageRank is not accurate, and that BFS is stronger, we do not compare our approaches with PageRank. We also exclude from our study CTR-based crawling due their reliance on historical data, and Depth-First Search as it generally underperforms BFS~\cite{boldi2004paradox}.
    For significance testing, we use a two-tailed paired Student's t-test with $p=0.01$ to compare each policy independently w.r.t. the baseline in terms of maxNDCG and nDCG@10, and a two-tailed Z-test for proportions with $p=0.01$ for comparisons in HR.
    
    The source code to reproduce our experiments is publicly available on Github\footnote{\url{https://github.com/fpezzuti/neural_crawling}}.
\begin{figure*}[t!]
    \includegraphics[width=1\linewidth]{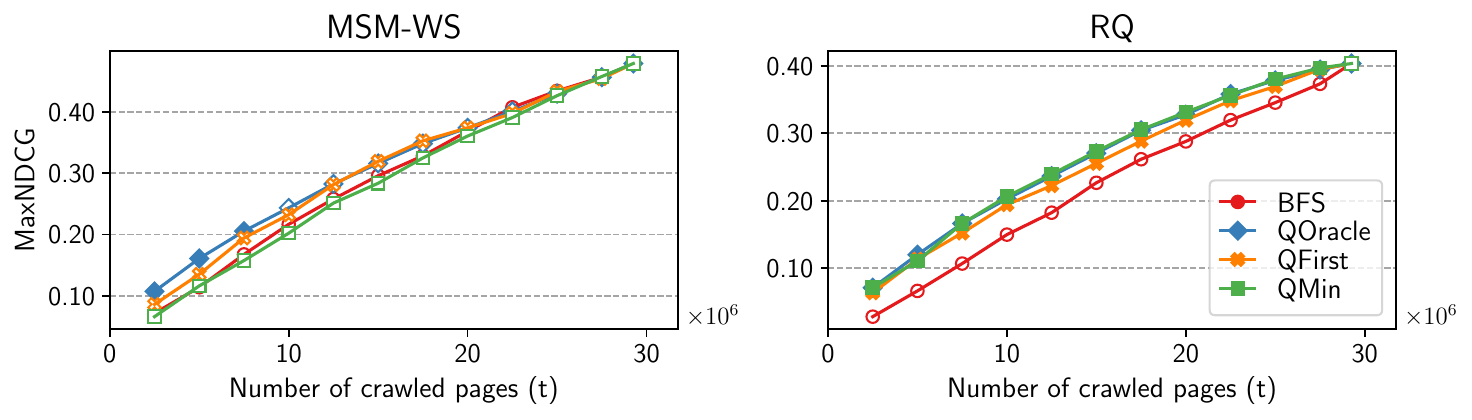}
    \caption{Assessment of neural crawling policies and \bfs\ in terms of maxNDCG. MSM-WS (left) and RQ (right). Statistically significant differences w.r.t. \bfs\ are denoted with \textit{filled circles} ($p=0.01$).}
    \label{fig:maxndcg}
    \Description[Neural crawlers are more effective than \bfs\ in terms of maxNDCG]{Both for MSM-WS and RQ, our neural crawling policies are more effective than \bfs\ in terms of maxNDCG during the first iterations of the crawl.}
\end{figure*}
\begin{figure*}[t!]
    \includegraphics[width=1\linewidth]{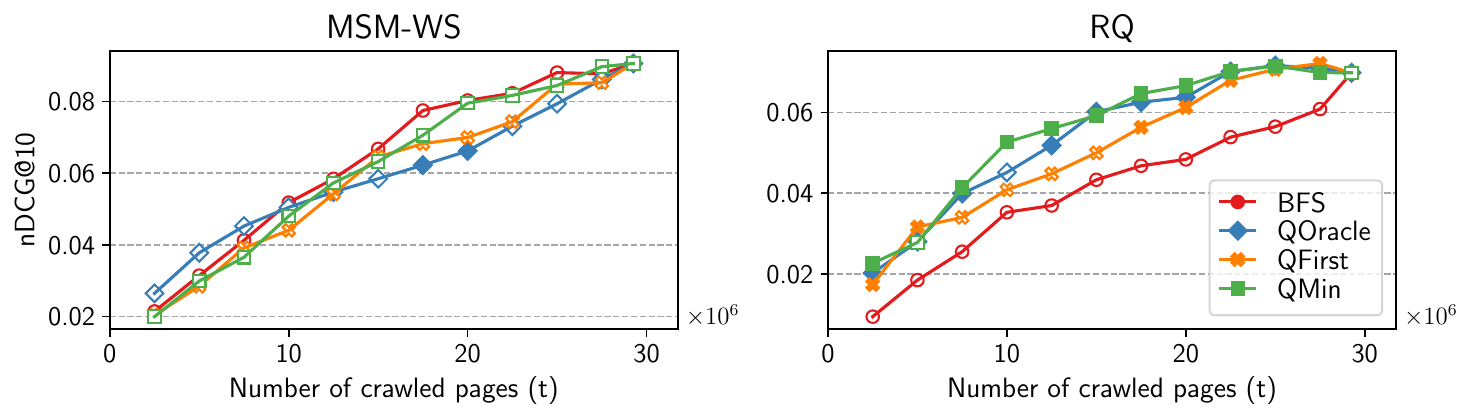}
    \caption{Assessment of neural crawling policies and \bfs\ in terms of nDCG@10 ranking effectiveness. MSM-WS (left) and RQ (right). Statistically significant differences w.r.t. \bfs\ are denoted with \textit{filled circles} ($p=0.01$).}
    \label{fig:reranking-ndcg-all}
\end{figure*}

\section{Results}
In this section, we answer our research questions and we extensively discuss our findings.

\subsection{RQ1. Crawling Effectiveness}
Firstly, we investigate whether leveraging neural policies during the crawl, the ability of a crawler in early fetching relevant pages can be improved compared to \bfs.
To tackle this, in Figures~\ref{fig:hr-all} and~\ref{fig:maxndcg} we report the HR and the maxNDCG at different time instants for our oracle neural policy, our two proposed practical neural policies, and \bfs, on MSM-WS and RQ.
From Figure~\ref{fig:hr-all}, we note that, in terms of HR, our \qoracle\ initially exhibits superior performance w.r.t. all other policies in surfacing pages relevant to keyword queries in MSM-WS, and is immediately followed by \qfirst. On the RQ query set, containing natural language queries, all our neural crawlers substantially outperform \bfs, and interestingly \qmin\ attains almost the same HR as \qoracle. The results shown in Figure~\ref{fig:maxndcg} indicate that on MSM-WS our \qoracle\ and \qfirst\ policies yield to a small increase in maxNDCG compared to \bfs\ and \qmin, which perform similarly. Meanwhile, they also indicate that on natural language queries from RQ, all neural crawling policies consistently outperform \bfs\ throughout the entire crawl. 
On RQ1, we conclude that a neural crawling policy yields to an increase in crawling effectiveness both for keyword queries and natural language queries, particularly remarkable with the latter.

\subsection{RQ2. Ranking Effectiveness}
Secondly, we investigate whether the effectiveness of a retriever can be increased exploiting our proposed neural crawling policies during the crawl, particularly in delivering early relevant results to challenging natural language queries.
To address this, in Figure~\ref{fig:reranking-ndcg-all} we report, at different time instants, the nDCG@10 of a \textit{BM25 $\gg$ MonoELECTRA} IR pipeline. We note that, throughout the entire crawl, all the three crawlers based on neural policies outperform \bfs on natural language queries from RQ while being almost on par with \bfs\ on keyword queries from MSM-WS.
In particular, on natural language queries, \qmin\ is slightly more effective than \qoracle, and both are substantially more effective than \qfirst. To conclude on RQ2, neural crawling policies can yield remarkable benefits also in terms of retrieval effectiveness for complex natural language queries.

\subsection{Effectiveness Trade-off}
Next, we discuss all neural policies in terms of overall effectiveness.
As expected, \qoracle\ achieves highest crawling and search effectiveness except on MSM-WS during the last stages of the crawl, where it performs the worst in terms of nDCG@10.
This suggests that while \qoracle\ effectively collects relevant pages for keyword queries, and excels at initially ranking them, its search performance deteriorates as the crawl proceeds, suggesting that the indexed corpus becomes filled of distracting pages for keyword matching.
Interestingly, \qmin, exhibits comparable performance to \qoracle\ on RQ, while being unexpectedly on par with \bfs\ on MSM-WS. This suggests that while most pages valuable for natural language search can be easily reached by postponing the exploration of low-quality links, some pages valuable for keyword-oriented search may only be reachable through low-quality links.
Hence, reluctance in following these links may hamper the discovery of valuable Web pages located deeper in the Web graph. 
Meanwhile, \qfirst, despite being the simplest neural crawling policy, significantly outperforms \bfs\ on both query sets, and achieves competitive performance w.r.t. the other methods without introducing excessive overhead.
Unlike \qoracle\ and \qmin, which rely on a greedier prioritisation and favour exploitation, \qfirst\ is more exploration-oriented as it relies on noisier estimates. As a result, it has higher chances of discovering valuable pages only reachable throughout local minima.

\subsection{RQ3. Crawling Efficiency}
Lastly, we compare the efficiency of crawlers guided by our proposed neural policies, in terms of mean speedup throughout the crawl.
From Table~\ref{tab:speedup}, we observe that on the query set of natural language queries, all our three crawling approaches outperform the \bfs\ baseline.
Notably, \qmin\ achieves a mean speedup of $1.6\times$ over \bfs, meaning that to collect the same number of relevant pages, \bfs\ needs to crawl and store approximately $60\%$ more pages than \qmin, and needs to spend $60\%$ more time crawling.
On MSM-WS, which contains keyword queries, the mean speedup achieved by our neural crawlers is generally lower. While our \qfirst\ approach is $1.1\times$ more efficient than \bfs, \qmin\ is approximately $4\%$ less efficient than \bfs --- however, the inefficiency is marginal. 
In summary, in addition to improving crawling effectiveness and downstream retrieval effectiveness, our proposed neural policies consistently enhance time and resource efficiencies throughout the crawl.

\begin{table}[t!]
     \centering
      \caption{Mean speedup throughout the crawl and relative to \bfs, achieved by crawling with our proposed neural prioritisation strategies.}
     \label{tab:speedup}
    \begin{tabular}{@{}cccc@{}}
        \toprule
        & \qoracle & \qfirst & \qmin \\
        \midrule
         MSM-WS & $1.302$ & $1.100$ & $0.967$ \\
         RQ     & $1.514$ & $1.405$ & $1.601$ \\
        \bottomrule
    \end{tabular}
 \end{table}

\section{Conclusions}
    In this paper, we introduced a novel approach to Web crawling, rethinking traditional methods and challenging the conventional assumptions to better align crawlers to the recent shift towards natural language search.
    
    Specifically, we investigated if leveraging a neural quality estimator to drive the crawling process in prioritising the early crawl of high-quality pages can improve the early effectiveness of a crawler and of a retriever, particularly for natural language queries.
    We proposed two neural policies for crawling and we compared them with an oracle policy and BFS. Our findings reveal that we can markedly improve both the effectiveness of the crawler and of the retriever of a Web search system by leveraging a neural quality estimator during the crawl, and our approach is especially impactful for natural language queries.
    Moreover, our experiments show that by guiding the crawl using our proposed approaches, we can also improve efficiency, particularly for natural language queries.
    
    While our results show the promise of our approach, we recognise several limitations of this work that open up meaningful directions for future research.
    First, our experiments were conducted in a controlled, simulated setting; the effectiveness of our approach has not been validated in real-world, multi-threaded environments, which are typically subject to practical constraints like politeness policies, host reachability issues, and others.
    Second, further investigation is needed to better understand the potential biases introduced by neural quality estimators, particularly in terms of fairness and transparency.
    Third, we acknowledge that our proposed policies may be vulnerable to adversarial manipulation, and their robustness to such attacks should has yet to be explored.
    Lastly, we leave for future work experiments on other Web corpora and query sets, as well as experiments with more advanced policies and additional baseline comparisons.
    
    In conclusion, this work rethinks Web crawling and sets the stage for a new generation of crawlers with embedded semantic understanding.
    These second-generation crawlers are, in principle, not only more aligned with modern natural language search, but also more effective, scalable, adaptable, and flexible.
    We view this contribution as both an innovative shift in perspective and a foundation upon which future research can build.

\begin{acks}
This work was partially supported by the Spoke ``FutureHPC \& BigData'' of the ICSC – Centro Nazionale di Ricerca in High-Performance Computing, Big Data and Quantum Computing funded by the Italian Government, the FoReLab and CrossLab projects (Departments of Excellence), the NEREO PRIN project funded by the Italian Ministry of Education and Research and European Union - Next Generation EU (M4C1 CUP 2022AEF-HAZ), and the FUN project (SGA 2024FSTPC2PN30) funded by the OpenWebSearch.eu project (GA 101070014).
\end{acks}

\balance

\bibliographystyle{ACM-Reference-Format}
\bibliography{bibliography}


\end{document}